\documentclass[a4paper]{jpconf}
\usepackage{graphicx}
\begin{document}

\title{Comparative study of the inclusive asymmetries induced by polarized protons and antiprotons at 16 GeV/$c$ at the U--70 accelerator}

\author{V A Okorokov$^{1}$, V V Abramov$^{2}$, A A Bogdanov$^{1}$, M A Chetvertkov$^{3}$, V A Chetvertkova$^{4}$,
V V Mochalov$^{1,2}$, S B Nurushev$^{1,2}$, \\ M B Nurusheva$^{1}$,V L
Rykov$^{1}$, P A Semenov$^{1,2}$, M N Strikhanov$^{1}$ and A N Vasiliev$^{1,2}$}

\address{$^{1}$National Research Nuclear University MEPhI (Moscow Engineering Physics Institute),
Kashirskoe highway 31, 115409 Moscow, Russia}

\address{$^{2}$National Research Center "Kurchatov Institute" -- Institute for High Energy Physics, Protvino,
Moscow region, 142281, Russia}

\address{$^{3}$MD Anderson Cancer Center, Department of Radiation Physics, 1515 Holcombe Blvd, Unit 1420,
Houston, TX 77030, USA}

\address{$^{4}$GSI Helmholtzzentrum f$\ddot{\mbox{u}}$r Schwerionenforschung GmbH, Plankstrasse 1, 64291, Darmstadt, Germany}

\ead{VAOkorokov@mephi.ru}

\begin{abstract}
The only comparative study of the inclusive pion single-spin
asymmetries produced in the interactions of the polarized protons
and antiprotons in collisions with unpolarized proton was carried
out at E--704 experiment. Significant asymmetries were found at
large $x_{F}$ and middle $p_{T}$, $\pi^{+}$ and $\pi^{0}$
asymmetries have positive signs while $\pi^{-}$ has negative one
in the $p^{\uparrow}+p$ collisions, while in the
$\bar{p}^{\uparrow}+p$ interactions the $\pi^{-}$ and $\pi^{0}$
asymmetries have positive signs while $\pi^{+}$ has negative sign.
Similar experimental study can be done in the SPASCHARM experiment
at U--70 accelerator at IHEP for various secondary particles with
the use of 16 GeV polarized proton and antiproton beams.

\end{abstract}

\section{Introduction}\label{sec:1}
Inclusive reactions measured with polarized (anti)proton beam give
insight into the spin dependence of the underlying partonic
processes and add new input regarding the problem of the spin
structure of polarized protons. Significant polarization effects
appear already at relatively low values of the transverse momentum $p_{T}$ ($p_{T} \sim 1.0$ GeV/$c$), where perturbative
quantum chromodynamics (pQCD) is not expected to be applicable.
Therefore the new experimental measurements with high statistics
as well as phenomenological studies are essential for future
progress in quantitative description of spin observables.

The following inclusive reactions are used for study of
single-spin asymmetries for pions
\begin{equation}
p^{\uparrow}+p \to \pi^{\pm,0}+X, \label{eq:1}
\end{equation}
\begin{equation}
\bar{p}^{\uparrow}+p \to \pi^{\pm,0}+X. \label{eq:2}
\end{equation}
The analyzing power $A_{N}$ is the physics observable under
consideration here. The $A_{N}$ is deduced from the measured
yields of pions produced in a well defined azimuthal angular
interval around the beam axis using vertically polarized $p$
($\bar{p}$) of both polarization signs:
\begin{equation}
A_{N}(z)=\frac{\textstyle 1}{\textstyle P_{B}\langle \cos \phi
\rangle}\frac{\textstyle
N_{\uparrow}(z,\phi)-N_{\downarrow}(z,\phi)}{\textstyle
N_{\uparrow}(z,\phi)+N_{\downarrow}(z,\phi)}. \label{eq:3}
\end{equation}
Here $z \equiv (p_{T},x_{F},\sqrt{s})$ -- a set of kinematic variables, $p_{T}$ is the pion transverse
momentum, $x_{F}=2p_{L}/\sqrt{s}$ -- Feynman variable for pion
with longitudinal momentum $p_{L}$ at the collision energy
$\sqrt{s}$, $P_{B}$ is the beam polarization, and $\phi$ is the
azimuthal angle between the beam polarization axis directed upward
and the normal to the production plane. $N_{\uparrow}$
($N_{\downarrow}$) is the number of pions produced for positive
(negative) spin orientation of the beam (anti)protons at the
target, normalized to the corresponding beam flux. The reactions
(\ref{eq:1}) and (\ref{eq:2}) were studied at $p=200$ GeV/$c$
($\sqrt{s}=19.4$ GeV) with the E--704 setup
\cite{PLB-264-464-1991} and here it is suggested to continue such a study in deeply
non-perturbative region at $p=16$ GeV/$c$ ($\sqrt{s}=5.64$ GeV) in the SPASCHARM project \cite{arXiv-0712.2691-2007,PPN-44-930-2013,CDR-draft-13082017}.

\section{The E--704 results for secondary pions} \label{sec:2}
The magnitude of $A_{N}$ increases for both $\pi^{+}$ and
$\pi^{-}$ particles with $x_{F}$ for $p^{\uparrow}$ beam, but the
sign of $A_{N}$ is negative for the $\pi^{-}$ data. The values of
$A_{N}$ are large for high values of $x_{F}$, up to $0.29 \pm
0.09$ for $\pi^{+}$ and down to $–0.37 \pm 0.07$ for $\pi^{-}$.
Detailed analysis of data at $p=200$ GeV/$c$ shown a threshold
effect in which $A_{N}$ increases dramatically above $p_{T}=0.7$
GeV/$c$ \cite{PLB-264-464-1991}. The E--704 experimental results
obtained for $A_{N}(x_{F})$ above and below $p_{T}$ threshold
indicated that the increase of $A_{N}$ is primarily $x_{F}$ effect
above a $p_{T}$ threshold. The situation is similar for study of the $A_{N}(x_{F})$
with $\bar{p}^{\uparrow}$ beam with taken into account the change of signs of electric charge for the beam and secondary particles
under consideration. Furthermore the $p_{T}$ dependence was
obtained for $A_{N}$ for charged pion production in inclusive
reaction (\ref{eq:2}) with $\bar{p}^{\uparrow}$
beam \cite{PRL-77-2626-1996}. As example, figure \ref{fig:2-1}
shows the dependence of $A_{N}$ on $x_{F}$ (left) and $p_{T}$
(right) for secondary charged pions in (\ref{eq:2}). In this case
the E--704 data exhibit an almost mirror symmetric dependence in
$x_{F}$. The analyzing power for $\pi^{-}$ production increases
from 0.0  to about +0.25 with increasing $x_{F}$ above $p_{T} \sim
0.5$ GeV/$c$ while, for $\pi^{+}$ production, $A_{N}$ decreases
from 0.0 to about -0.35 with increasing $x_{F}$ above the same
$p_{T}$. The E--704 results with $\bar{p}^{\uparrow}$ show a
threshold effect about $p_{T} \sim 0.5$ GeV/$c$, above which
$A_{N}$ increases in magnitude for both $\pi^{+}$ and $\pi^{-}$
and for transverse momenta below this $p_{T}$ value, $A_{N}$ is
significantly smaller and compatible with zero
\cite{PRL-77-2626-1996}. The threshold effect was also confirmed
by the results of additional analysis for $p_{T} \geq 0.5$
GeV/$c$. Therefore the E--704 experimental results for
(\ref{eq:2}) show the magnitude of $A_{N}$ increases for both types
of charged pions with increasing $x_{F}$, but the sign of $A_{N}$
is positive for the $\pi^{-}$ data and negative for $\pi^{+}$ data
above the same $p_{T} \sim 0.5$ GeV/$c$. It appears that $A_{N}$
depends primarily on $x_{F}$ namely, and reaches large values
above the $p_{T}$ threshold of about 0.5 GeV/$c$ as well as for
inclusive reaction (\ref{eq:1}) with charged pions at some higher
threshold $p_{T}$.

\begin{figure}[h!]
\begin{center}
\includegraphics[width=12.5cm,height=9.0cm]{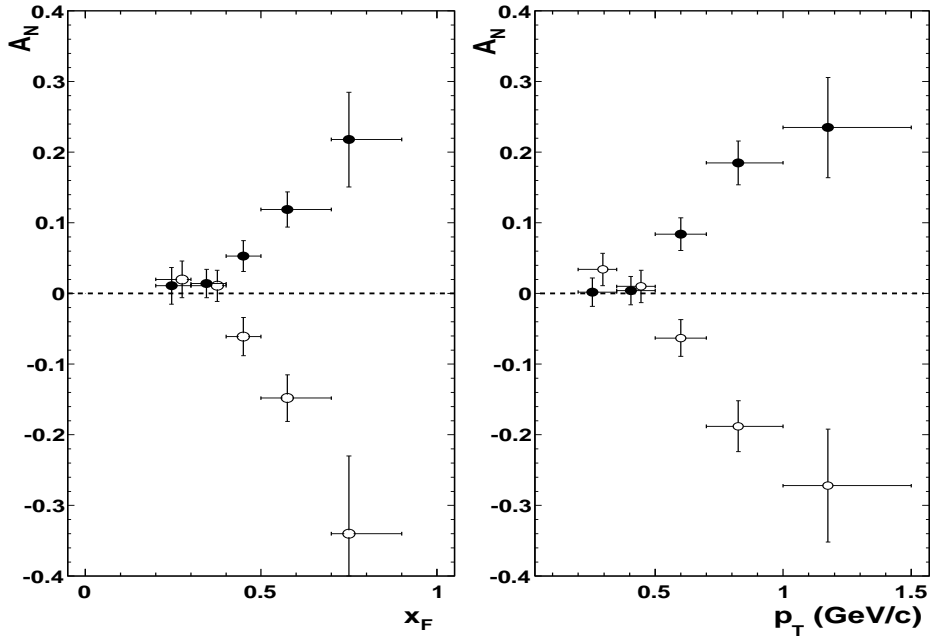}
\end{center}
\caption{\label{fig:2-1}$A_{N}$ data for $\bar{p}^{\uparrow}$ beam
as a function of $x_{F}$ integrated over $p_{T}$  in the range
$0.2-1.5$ GeV/$c$ (left) and depends on $p_{T}$ in the $x_{F}$
range of $0.2-0.9$ (right) for charged pions. Experimental points
for $\pi^{-}$ are shown by solid symbols, for $\pi^{+}$ -- by open
ones. For clarity, first two $\pi^{-}$ ($\pi^{+}$) data points are
slightly shifted on left (right) in each panel. Data are taken
from \cite{PRL-77-2626-1996}.}\label{fig:2-1}
\end{figure}

Therefore $A_{N}$ seems similar (in the sense of behavior and magnitude)
for $p^{\uparrow}$ / $\bar{p}^{\uparrow}$ beams in the case of
opposite charges for pions, i.e. the reaction (\ref{eq:1}) with
proton beam and inclusive $\pi^{+}$ the $A_{N}$ appears to be
similar to that in reaction (\ref{eq:2}) with antiproton beam and
inclusive $\pi^{-}$ and vise versa. One can note that the models
based on non-perturbative approaches, such as a soft pion exchange
mechanism \cite{PRL-68-907-1992}, resonance-decay interference
between real and virtual channels \cite{PLB-296-251-1992} and
rotating constituents in the polarized (anti)proton
\cite{PRL-70-1751-1993} appear to be in good qualitative agreement
with the features of the data on the pion production asymmetry
measured with the both polarized protons and antiprotons.

The kinematic dependencies of $A_{N}$ on $x_{F}$, $p_{T}$ were
also studied for the $\pi^{0}$ production in the inclusive reactions
(\ref{eq:1}) and (\ref{eq:2}) in E--704 experiment
\cite{PLB-264-464-1991,PRL-61-1918-1988,PLB-261-201-1991,PRD-53-4747-1996}.
For the $\pi^{0}$ data obtained with $p^{\uparrow}$ beam,
$A_{N}(x_{F})$ has the same sign as for $\pi^{+}$ data and is
about half as large \cite{PLB-264-464-1991,PLB-261-201-1991}. The
similar situation is observed for $A_{N}(x_{F})$ in the
$\bar{p}^{\uparrow}+p$ inclusive reaction for comparison $\pi^{0}$
results with $\pi^{-}$ analyzing power
\cite{PRL-77-2626-1996,PLB-261-201-1991}. Thus the $\pi^{0}$
productions by polarized $p^{\uparrow}$ and $\bar{p}^{\uparrow}$
are related by charge conjugation of the beam and the produced
particle. The measured asymmetries have the same sign and similar
$x_{F}$  dependence \cite{PLB-261-201-1991}. At large $x_{F}$,
there is an indication that the magnitude of $A_{N}$ for $\pi^{0}$
production by incident antiprotons is less than for incident
protons. This would mean that the interactions involve
constituents other than gluons and quark-antiquark pairs in the
target proton. The $A_{N}$ is observed to be zero for single-spin
inclusive $\pi^{0}$ production in $p^{\uparrow}+p$ and
$\bar{p}^{\uparrow}+p$ inclusive reactions in the $1 < p_{T} < 3$
GeV/$c$ region within a statistical accuracy
\cite{PRD-53-4747-1996}. But it should be noted that in this case
the amount of data for studying (\ref{eq:2}) interactions was an
order of magnitude less than that for (\ref{eq:1}) interactions.
In general in perturbative QCD single-spin transverse asymmetries
are expected to be practically zero. Thus this expectation in the
$1 < p_{T} < 3$ GeV/$c$ region is confirmed by the data from the
E--704 experiment, if perturbative QCD is applicable to these
$p_{T}$ values at beam momentum $p=200$ GeV/$c$. Moreover the
experimental errors are large for $A_{N}$ at $p_{T} > 2.5$ GeV/$c$
\cite{PRD-53-4747-1996}. Therefore new experimental data with high
statistics seem important for verification of some predictions of
the QCD.

\section{The SPASCHARM project} \label{sec:3}
The SPASCHARM ({\bf SP}in {\bf A}symmetry in {\bf CHARM}onia) is
the project for world-class research works in fixed target mode
for high energy spin physics \cite{arXiv-0712.2691-2007,PPN-44-930-2013,CDR-draft-13082017}. The main goals of the SPASCHARM
project are the studies of the (i) spin structure of the nucleon
and (ii) possible spin dependence of the strong interaction for
matter and antimatter with help of the systematic physical
analysis for a wide set of hadronic reactions and secondary
particles. The same name experimental setup is the core part of
the SPASCHARM project. It is suggested to have two stages of the SPASCHARM
project: first of all, studies with unpolarized proton beams using a polarized target and
second phase of the project is the using of polarized beams.

The polarized $p^{\uparrow}$ ($\bar{p}^{\uparrow}$) beam is
obtained by selecting $p^{\uparrow}$ ($\bar{p}^{\uparrow}$) from
the weak decay of $\Lambda$ ($\bar{\Lambda}$) produced in a primary target
by extracted proton beam. This method is used for both the E--704
experiment and the SPASCHARM setup. The main features are shown
in table \ref{table:1} for U--70 and Tevatron beams. As seen from
table \ref{table:1} there are some advantages of the expectations
for U--70 (for instance, intensities for primary and polarized
proton beams) with respect to corresponding Tevatron beam
parameters. It should be noted the U--70 allows the study of
single-spin asymmetry in deeply non-pertubative region for some
range of $\sqrt{s}$.

\begin{center}
\begin{table}[h]
\caption{\label{table:1} Some main characteristics for U--70 and Tevatron beams.} \centering
\begin{tabular}{@{}*{4}{c}}
\br \multicolumn{1}{c}{}  & \multicolumn{1}{c}{Beam parameter} &
\multicolumn{1}{c}{U--70} &
\multicolumn{1}{c}{Tevatron}\\
\mr
1 & primary proton beam, $p$ (Gev/$c$) & 50--60 & 800 \\
2 & primary beam intensity, c$^{-1}$  & $\sim 2 \times 10^{12}$ & $1.5 \times 10^{11}$ \\
3 & polarized beam, $p$ (Gev/$c$) & 15--45$^{a}$ & $185 \pm 17$ \\
4$^{b}$ & beam intensity at the target, c$^{-1}$ & $(0.9-6.8) \times 10^{6}$ & $1.5 \times 10^{6}$ \\
  &  & $(0.8-4.0) \times 10^{4}$ & $1.5 \times 10^{5}$ \\
5$^{b}$ & beam polarization & $\pm (0.45 \pm 0.05)$ & $\pm (0.40 \pm 0.12)$ \\
  &  & --//-- & $\pm (0.45 \pm 0.03)$ \\
\br
\multicolumn{4}{l}{\footnotesize{$^{a}$The minimum (maximum) relative uncertainty for beam momentum $\delta p$ is 4.5\% (11\%)}}\\
\multicolumn{4}{l}{\footnotesize{~~for $p=15$ GeV/$c$ and 3.0\% (9.0\%) for $p=45$ GeV/$c$.}}\\
\multicolumn{4}{l}{\footnotesize{$^{b}$The first / second line
corresponds to the $p^{\uparrow}$ / $\bar{p}^{\uparrow}$ beam.}}\\
\end{tabular}
\end{table}
\end{center}

\begin{figure}[h!]
\begin{center}
\includegraphics[width=12.5cm]{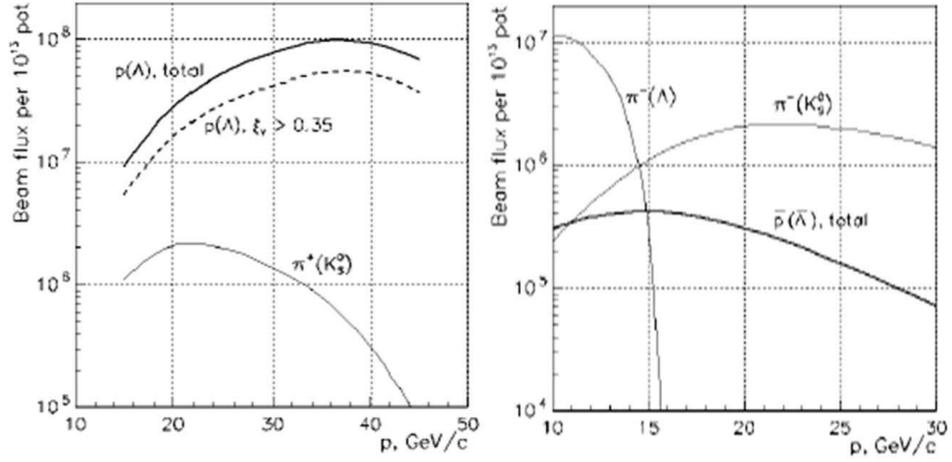}
\end{center}
\caption{\label{fig:3-1}Intensity for beam of $p^{\uparrow}$ (left) and $\bar{p}^{\uparrow}$ (right) along with pion background of appropriate sign of electric charge for maximum $\delta p$ shown in table \ref{table:1}. The quantity is calculated for $10^{13}$ primary protons with energy 60 GeV, where $\xi_{y}$ is the transverse polarization averaged over ensemble. Data are taken from \cite{CDR-draft-13082017}.}
\end{figure}

Dependence of polarized beam intensity on momentum is shown in
figure \ref{fig:3-1} for $p$ (left) and $\bar{p}$ (right). The
difference in intensities of $p^{\uparrow}$ and
$\bar{p}^{\uparrow}$ beams  increases dramatically with a growth
of beam momentum. As seen for polarized proton beam the
contribution of $\pi^{+}$ from neutral kaon decays is small,
furthermore this contribution decreases rapidly with beam momentum.
Therefore the U--70 allows the polarized proton beam with good
quality (purity). Figure \ref{fig:3-1} (right) shows that the beam
of $\bar{p}^{\uparrow}$ with $p=16$ GeV/$c$ seems optimal in terms
of intensity and background conditions. The number of $\pi^{-}$ is
approximately 3 times higher than the intensity of the antiproton
beam. The separation of antiprotons at this level of background is
quite possible with the help of Cherenkov beam counters.

\begin{figure}[h!]
\begin{center}
\includegraphics[width=16.0cm]{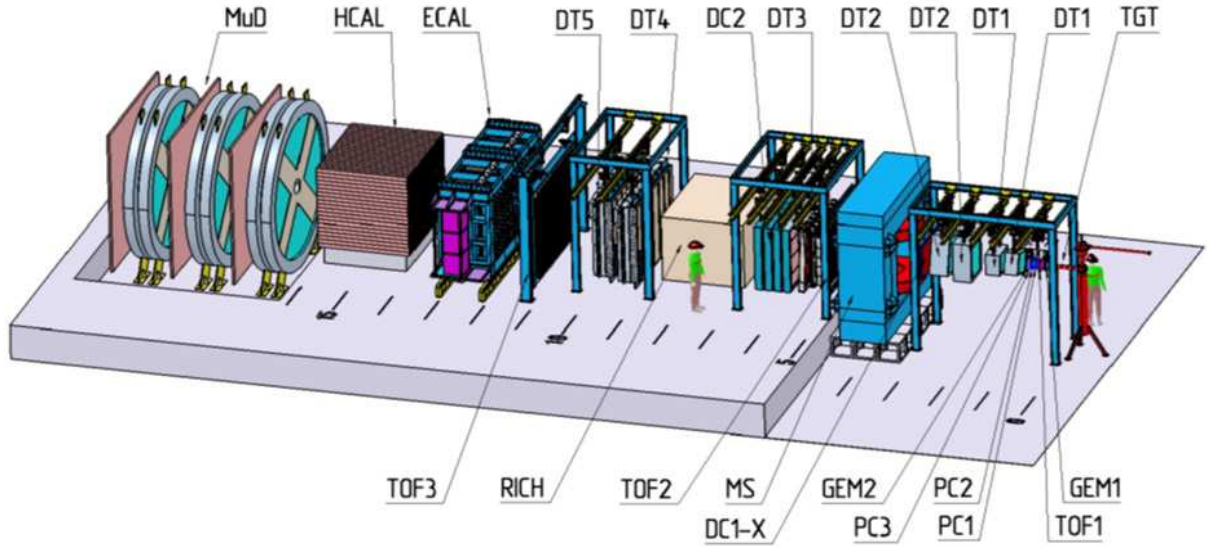}
\end{center}
\caption{\label{fig:3-2}Schematic 3D view of the SPASCHARM setup. Here liquid hydrogen target (TGT) is shown, DT1--5 -- a sets of thin-walled drift tubes, other subsystems are described in the text. Distances are shown in meters along the beam direction with respect to the center of the target.}
\end{figure}

\begin{figure}[h!]
\begin{center}
\includegraphics[width=12.5cm]{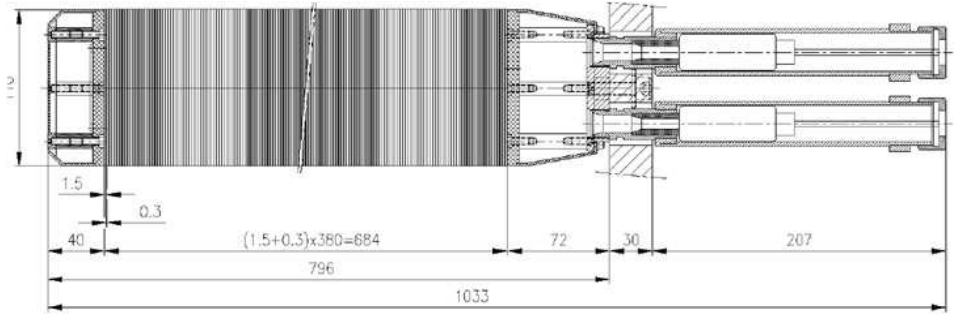}
\end{center}
\caption{\label{fig:3-3}Scheme of supermodul for ECAL ("shashlik" consists from layers of $\mbox{Pb}$-absorber and scintillator). Picture is taken from \cite{CDR-draft-13082017}.}
\end{figure}

The SPASCHARM detector is an open geometry setup with good particle identification and relatively large acceptance. Schematic view of the SPASCHARM is shown in figure \ref{fig:3-2}. The SPASCHARM setup consists of the following main subsystems \cite{arXiv-0712.2691-2007,CDR-draft-13082017}: \\
-- various targets (liquid hydrogen, nuclear from $\mbox{Be}$ up to $\mbox{Pb}$; polarized / unpolarized), \\
-- spectrometer for registration of charged particles (magnet, GEM detector, multiwire proportional chambers -- PC and drift ones -- DC),\\
-- electromagnetic calorimeter (ECAL) is shown in figure \ref{fig:3-3} and it is made by "shashlik" technology which is well established and used successfully, for instance, during preparation the PANDA experiment, \\
-- hadronic calorimeter (HCAL),\\
-- set of detectors for particle identification and multiplicity
measurement (ring image Cherenkov detector -- RICH, muon
spectrometer -- MuD and time-of-light system -- TOF).

Subsystems of the SPASCHARM detector are designed for high quality
particle identification. According to the Conceptual design report
(CDR) \cite{CDR-draft-13082017} the SPASCHARM will provide registration of charged
particles from 0.5 up to 50 GeV/$c$ in full azimuth $2\pi$,
registration of $\gamma$ from 0.2 up to 50 GeV/$c$ with the same
angular range as for charged particles, identification of
charged particles for momentum domain 1--20 GeV/$c$, registration
of decay particles from hyperons with future reconstruction of
parent particle, high coordinate resolution for beam particles,
especially, for elastic scattering study
\cite{CDR-draft-13082017}. Furthermore there are following special
requests for charmonium study: momentum resolution is $\sim 2\%$
at 10 GeV/$c$ and energy resolution for electromagnetic
calorimeter is $3\% / \sqrt{E}$. Therefore there are some
advantages of the SPASCHARM project regarding of earlier
experiments, for instance,  E--704. In particular, addition of new
detectors (GEM, MDC, high quality EMC etc.) allows the increase of
statistics significantly compared to the previous experiments. Due
to $2\pi$ acceptance on azimuthal angle of the SPASCHARM setup one
can expect that the systematic errors in $A_{N}$ will be small.
With the use of polarized proton beam at SPASCHARM a precision
measurement of $A_{N}$ for inclusive production in the transverse
polarized beam fragmentation region in a wide ($x_{F}$,
$p_{T}$)-region will be worthwhile \cite{arXiv-0712.2691-2007}. One can expect that the kinematic ranges $0.2 < x_{F} < 1.0$ and $0.5 < p_{T} < 3.5$ GeV/$c$ will be covered. Also the estimations will be obtained for accuracy of the $A_{N}$ measurement in the reactions (\ref{eq:1}), (\ref{eq:2}) and for corresponding time of data collection for various accelerator parameters (beam intensity, run duration etc.).

Within SPASCHARM project the software is constructed for on-line
and off-line data analyses. The software is developed as
object-oriented environment and it is based on the ROOT package.
Also SPASCHARM software includes GEANT 3, 4 and some event
generators (PYTHIA, PLUTO etc.) for Monte-Carlo simulation of
hadronic interactions. At present the software environment
SpascharmRoot has been developed and partially implemented. The
SpascharmRoot allows the simulation, reconstruction, on- and
off-line analysis of experimental data. The environment is under
active development. The SPASCHARM software is permanently
improving, for instance,  the possibility is considered for using
of the GRID technology for distributed data analysis.

\section{Summary} \label{sec:4}
Analyzing power shows the significant magnitude for inclusive
pions above $p_{T} \sim 0.5$ GeV/$c$ but there is only one
measurement of the $A_{N}$ for inclusive pions with polarized
antiproton beam. At present phenomenological models describe the
experimental data at qualitative level only and precision of
experimental results do not allow the clear discrimination between
various models. Consequently the new high-precision measurements
seem important for better understanding of single-spin asymmetry
and more definite physics conclusions. Hopefully, the SPASCHARM
experiment will provide the high-statistics data which will shed
new light for single-spin pion asymmetry as well as in general for
spin structure of the proton.

\section*{Acknowledgments}

The work has been supported in part by the RFBR Grant No 16--02--00667
and by the NRNU MEPhI Academic Excellence Project (contract No 02.a03.21.0005, 27.08.2013).

\smallskip
\end{document}